\documentclass[12pt,preprint]{aastex}

\def\gta{\ifmmode {\mathbin{\lower 3pt\hbox   
    {$\,\rlap{\raise 5pt\hbox{$\char'076$}}\mathchar"7218\,$}}}
    \else {${\mathbin{\lower 3pt\hbox
    {$\rlap{\raise 5pt\hbox{$\char'076$}}\mathchar"7218\,$}}}
    $}\fi}
\def\lta{\ifmmode {\,\mathbin{\lower 3pt\hbox   
    {$\,\rlap{\raise 5pt\hbox{$\char'074$}}\mathchar"7218\,$}}}
    \else {${\mathbin{\lower 3pt\hbox
    {$\rlap{\raise 5pt\hbox{$\char'074$}}\mathchar"7218\,$}}}
    $}\fi}

\shorttitle{BURST OSCILLATION LIGHT CURVES OF XTE J1814-338}
\shortauthors{Bhattacharyya et al.}

\begin{document}

\title{CONSTRAINTS ON NEUTRON STAR PARAMETERS FROM BURST OSCILLATION
LIGHT CURVES OF THE ACCRETING MILLISECOND PULSAR XTE J1814-338}

\author{Sudip Bhattacharyya\altaffilmark{1}, Tod
E. Strohmayer\altaffilmark{2}, M. Coleman Miller\altaffilmark{1}, and
Craig B. Markwardt\altaffilmark{1,2}}

\altaffiltext{1}{Department of Astronomy, University of Maryland at
College Park,
College Park, MD 20742-2421; sudip@astro.umd.edu, miller@astro.umd.edu}

\altaffiltext{2}{Laboratory for High Energy Astrophysics, Goddard Space 
Flight Center, NASA, Greenbelt, MD 20771; stroh@clarence.gsfc.nasa.gov, 
craigm@milkyway.gsfc.nasa.gov}

\begin{abstract}

Detailed modeling of the millisecond brightness oscillations
during thermonuclear bursts from low mass X-ray binaries can
provide important information about neutron star structure.
Until now the implementation of this idea has not been
entirely successful, largely because of the negligible
harmonic content in burst oscillation lightcurves. However,
the recent discovery of non-sinusoidal burst oscillation
lightcurves from the accreting millisecond pulsar XTE
J1814-338 has changed this situation. We, therefore, for the
first time, make use of this opportunity to constrain
neutron star parameters. In our detailed study
of the lightcurves of 22 bursts, we fit the burst
oscillation lightcurves with fully general relativistic
models that include light-bending and frame-dragging for
lightcurve calculation, and compute numerically the
structure of neutron stars using realistic equations of
state. We find that for our model and parameter grid values, 
at the 90\% confidence level, $Rc^2/GM > 4.2$ 
for the neutron star in XTE J1814-338. We also find
that the photons from the thermonuclear flash come
out through the layers of accreted matter under conditions
consistent with Thomson scattering, and show that the secondary companion
is a hydrogen burning main sequence star, with possible
bloating (probably due to X-ray heating). 

\end{abstract}


\section{Introduction} \label{sec: 1}

Nearly coherent brightness oscillations have been discovered with the
{\it Rossi} X-ray Timing Explorer ({\it RXTE}) during thermonuclear
X-ray bursts from more than a dozen neutron stars in low-mass X-ray
binaries (Strohmayer \& Bildsten 2003). 
The large modulation amplitudes at the onset of bursts, the time
evolution of the pulsed amplitude during the rise of bursts (Strohmayer,
Zhang, \& Swank 1997), 
the coherence of the oscillations (Smith, Morgan, \& Bradt 1997; 
Strohmayer \& Markwardt 1999; Muno et al. 2000), and the long term 
stability of the oscillation frequencies (Strohmayer et al. 1998)
led quickly to an interpretation in terms of a ``hot spot" on the surface 
of the star,
so that the observed flux is modulated by rotation (Strohmayer et al. 1996).  
This interpretation has been strongly supported by observations of the
transient accretion-powered pulsars SAX~J1808--3658 (Chakrabarty et al. 2003) 
and XTE~J1814--338 (Markwardt, Strohmayer, \& Swank 2003), 
in which the frequency of burst oscillations is
indeed extremely close to the spin frequency as inferred from
persistent oscillations.

Early on it was suggested that the intensity profiles of the pulse
lightcurves contain valuable information about the stellar mass and
radius (and hence about the equation of state of the dense matter in
the stellar core), as well as the surface emission properties and
system geometry (Strohmayer, Zhang, \& Swank 1997; 
Miller \& Lamb 1998; Weinberg, Miller,
\& Lamb 2001). The reason is that the lightcurves are
affected by general relativistic light deflection as well as special
relativistic beaming and aberration, which can be significant at the
$\sim 0.1-0.2$c linear rotational speeds implied for spin frequencies
of a few hundred Hertz. However, much of the required information is
encoded in the ratios of pulse amplitudes at different harmonics of
the spin frequency (although some constraints can be obtained if the
amplitude at the fundamental is particularly high; see Nath,
Strohmayer, \& Swank 2002). Until recently, no source had
harmonics detected definitively (for upper limits see Muno, \"{O}zel, \&
Chakrabarty 2002).

This situation has now changed. Analysis of XTE~J1814--338, the fifth
known accreting millisecond pulsar, shows a clear overtone in the
pulse profiles of many individual bursts, with an amplitude that can
be more than 0.25 times that of the fundamental (Strohmayer et al. 2003). 
In addition, the
314~Hz frequency of the fundamental oscillation is consistent with the
spin frequency seen from the persistent pulsations.  The existence of
harmonics and the confirmation of the basic hot spot picture make this
source very promising for detailed analysis and constraints on
stellar, geometrical, and emission properties.  We note that even more
precise information is in principle available from the extremely
accurately characterized lightcurves of the persistent pulsations
of millisecond pulsars,
which have several harmonics (e.g., see Cui, Morgan, \& Titarchuk 1998; 
Poutanen \& Gierli{\'n}ski 2003). However, the spectrum and profiles
of pulses during accretion are affected by complicated details of
shock physics and Comptonization. In contrast, thermonuclear burst
spectra are well fit by blackbodies, without any trace of a
high-energy tail related to Comptonization. They are also consistent
with coming from very near the stellar surface. For these reasons, the
pulses of XTE~J1814--338 are most promising for an initial
investigation.  Combined with analyses of surface spectral line profiles
(e.g., from EXO 0748-676; Cottam, Paerels, \& Mendez 2002), thermonuclear
X-ray bursts are key sources for constraining the state of matter
inside neutron stars.

Here we report our analysis of the {\it RXTE} PCA data on 22 
thermonuclear bursts from XTE~J1814--338.  In \S~2 we briefly describe 
our data extraction techniques.  In \S~3, we 
discuss our theoretical models, the fitting procedure and the method
of calculation of confidence intervals. We discuss our results in \S~4.
In particular, we show that for the two high-density equations of
state (EOS) that we analyze, there
is a preference for comparatively less compact neutron stars.
We also show that the emission pattern from points on the surface is
constrained moderately strongly and is close to that expected from 
the law of darkening for
passive radiative transport in a Thomson scattering atmosphere
(Chandrasekhar 1960).
In this paper we use geometric units, i.e., $G=c\equiv 1$.

\section{Data Analysis} \label{sec:2}

XTE J1814--338 was discovered in the Galactic bulge monitoring
campaign carried out with the {\it RXTE} PCA (Markwardt \& Swank
2003).  The system has a binary orbital period of 4.28 hours and the
neutron star has a spin frequency of 314.36 Hz (Markwardt et
al. 2004, in preparation). This is the widest binary among the
five known accreting millisecond pulsar systems. A total of 27 X-ray bursts
were observed during the extensive RXTE follow-up of the outburst. A
summary of basic burst properties can be found in Strohmayer et
al. (2003).  For the purposes of our modeling here we wished to
produce lightcurves with as high a statistical precision as possible,
while still maintaining energy resolved profiles. To do this we
elected to phase-align different bursts (while preserving any 
misalignments among the energy bands).
The energy channel boundaries were chosen so that the lightcurves in each
band would have roughly similar statistical qualities (ie. similar
numbers of total counts).  We used the channel boundaries; 0-3, 4-6,
7-10, 11-13 and 14-28 of the ``E\_125us\_64M\_0\_1s'' PCA event
mode. This corresponds to energy boundaries of $\approx$ 2 - 3.7 keV,
3.8 - 5.0 keV, 5.2 - 6.6 keV, 6.8 - 9.2 keV, and 9.4 - 23 keV.  To
co-add the bursts we found an offset phase for each burst which
maximized the $Z_2^2$ signal when adding the burst in question to the
initial, reference burst. This is basically the same procedure
described by Strohmayer \& Markwardt (1999).

Although the different burst profiles are qualitatively similar, and
hence adding them together is reasonable (see Figure 1 for added burst 
profile), we found that there is a
general trend for the harmonic strength to decrease with time.  That
is, bursts occurring later in the outburst seem to have somewhat
smaller ratios of the amplitude of the first overtone compared to that of the
fundamental.  This suggests that some secular change associated with
the accretion rate might be influencing the details of the pulse
profiles.  Therefore, to allow for such changes while still obtaining
good statistics, we added together the bursts in three groups
based on their times of occurrence (see Figures 2, 3 and 4 for
added burst profiles for three groups).  Since the overall structure
(ie. the mass and radius) of the neutron star must remain fixed
through the outburst, the secular changes in the pulse profile may
reflect small changes in the spot size, location or perhaps beaming
function, with time. We will say more about this shortly.

\section{Physical Effects and Calculations} \label{sec:3}

The basic picture we employ is one in which after leaving the photosphere, 
photons propagate freely in vacuum to the observer. This approach
ignores any scattering during the propagation.  If there were 
scattering by a hot plasma near the star, we would expect this to
leave its Comptonization imprint on the spectrum as an extended
power-law tail. The lack of such a signature (Strohmayer
et al. 2003) supports the assumption of free propagation.

Even with this simplification, if the emission pattern on the surface
and the angle-dependence of specific intensity from a given point on
the surface are arbitrary, then there are too many free parameters and
no meaningful constraints are possible.  Therefore, to make progress
in modeling, we adopt the following assumptions:

\begin{itemize}

\item The surface consists of a background of uniform intensity and
spectrum, plus a single hot spot that is a filled circle that emits
uniformly. This is the most popular simple emitting system for burst
oscillations. In a future paper we will consider models with
two hot spots.

\item For $R/M < 3.52$ and for Schwarzschild spacetime, an 
emitted photon is deflected by more than 180$^\circ$ (Pechenick, Ftaclas,
\& Cohen 1983). Therefore, to simplify numerical techniques 
by ensuring that no emitted photon is deflected by more
than 180$^\circ$, the minimum value of $R/M$ we consider is 3.6.

\end{itemize}

With these assumptions, we calculate lightcurves by tracing rays
backwards, i.e., from the observer to the stellar surface.  The paths
of these rays, and their specific intensities, carry the imprint of
several physical effects. Using an approach similar to that of 
\"Ozel \& Psaltis (2003), we consider: (1)~Doppler
boosts, (2)~special relativistic beaming, (3)~gravitational redshift,
(4)~light-bending, and (5)~frame-dragging.  We do not include the
effects of spin-induced stellar oblateness. These effects are only
second order in the stellar rotation and are thus small compared to
other uncertainties. For example, for all our EOS models, the polar to
equatorial radii ratio of the star is greater than 0.98 (for stellar
mass $M \ge 1.4 M_\odot$ and observed stellar spin frequency $\nu_{\rm *}
\sim 314$~Hz). We describe tests of our raytracing and lightcurve codes
in the Appendix.

To make our model lightcurves realistic, we calculate the relation
between $M$ and the stellar radius $R$, as well as $a/M$ (where, $a$
is the stellar angular momentum per unit mass) for a given $M$ and
$\nu_{\rm *}$ ($\sim 314$~Hz for XTE J1814-338), with a specified EOS
model.  We do this by computing numerically the stable structure of a
rapidly spinning neutron star, using the formalism given by Cook,
Shapiro, \& Teukolsky (1994) (for a detailed description, see also 
Bhattacharyya et al. 2000; Bhattacharyya 2002). 
For a particular neutron star EOS model and
assumed values for the stellar mass and spin rate, we solve Einstein's
field equations and the equation of hydrostatic equilibrium
self-consistently to obtain other stellar structure parameters
(radius, angular momentum, etc.). We use these output parameters in
our timing calculations.

In our detailed lightcurve calculation, we have five input parameters,
for a given EOS model and the known value of $\nu_{\rm *}$.  In our
study, the structure of the star is fixed by one unknown parameter,
which we choose to be (1)~the dimensionless stellar radius to mass
ratio $(R/M)$. The hot spot is specified by two parameters: (2)~the
$\theta$-position $\theta_{\rm c}$ of the center of the spot, and
(3)~the angular radius $\Delta \theta$ of the spot.  The emission from
a single point on the spot, as measured in the corotating frame, is
characterized by (4)~a parameter $n$ that gives a measure of the
beaming in the emitter's frame (corotating with the star), 
where the specific intensity as a function
of the angle $\psi$ (in the emitter's frame)
from the surface normal is $I(\psi) \propto
\cos^n(\psi)$. Finally, we have (5)~the inclination angle $i$ of the
observer as measured from the upper rotational pole.  
It is to be
noted, that at the initial phase of our model calculation, we consider
that the hot spot emits as a blackbody (as burst spectra can be fitted
well by blackbodies) with a temperature $kT = 2$~keV.  However,
later we allow the source spectrum to deviate from a blackbody (for
discussion about the spectrum and the justification of the assumed
temperature, see later in this section).

In our work, we consider two representative EOS models: 
A18 and A18$+\delta v+$UIX
(Akmal, Pandharipande, \& Ravenhall 1998). 
The first one is the Argonne $v_{18}$ model
(Wiringa, Stoks, \& Schiavilla 1995) of two-nucleon-interaction. This
model fits the Nijmegen database (Stoks et al. 1993) with
$\chi^2/N_{\rm data} \sim 1$, and hence is called ``modern". The model
A18$+\delta v+$UIX considers two additional physical effects: the
three-nucleon-interaction (Urbana IX (UIX) model; Pudliner et
al. 1995) and the effect of relativistic boost corrections to the A18
interaction. The EOS model A18
is soft, i.e., the maximum non-rotating mass it supports is
small $(\sim 1.67 M_\odot)$. On the other hand, the EOS model A18$+\delta
v+$UIX is hard (maximum supported non-rotating mass is $\sim 2.2 M_\odot)$. 
Figure 2 shows the stable stellar mass vs. radius curves for
these EOS models (for stellar spin frequency $\sim 314$~Hz).
Although, there are many other EOS models available in the literature, 
our chosen models span a representative range.

In order to compute the energy dependent flux, we trace back the paths
of the photons from the observer to the hot spot, using the Kerr
spacetime. We solve the geodesic equations numerically using a fifth order
Runge-Kutta method with adaptive stepsize control. The accuracy
of the results is tested and described in the appendix. Our ray-tracing
method is similar to that of Bhattacharyya, Bhattacharya, \& Thampan (2001),
except that here we use the Kerr spacetime, instead of the
correct spacetime around a rapidly rotating neutron star, for the sake of
simplicity. It is unlikely that the errors introduced 
by this approximation will have any detectable effect on the lightcurves
for the stellar spin frequency we consider (i.e., $\sim 314$ Hz).
Our method has two major differences from the approach of, e.g., 
Braje, Romani, \& Rauch (2000). (1) We 
solve the geodesic equations of motion of a photon from the observer
to the surface of the neutron star, instead of from the surface to the
observer. (2) Instead of doing elliptic integral reduction
of the geodesics (as was done by Braje et al. 2000), 
we solve the equations of motion numerically, for easier generalization to
numerical spacetimes.

The ray-tracing enables us to get the boundary of the
image of the source at the observer's sky.  We then calculate the
observed flux by integrating the observed energy dependent specific
intensity inside the boundary of this image (for a detailed
description, see Bhattacharyya et al. 2001). The
model lightcurve is calculated by repeating the same procedure for
many spots at different $\phi$-positions (but the same $\theta$-position) 
on the surface of the star. 
The actual phase points of the lightcurve
are calculated from these $\phi$-positions, the stellar spin frequency,
and the time delay consideration. The time delay is because
photons emitted at different points on the stellar surface
take different times to reach the observer. 

In this paper, we compare our models with the data of 22 bursts.
Based on burst occurrence time and the harmonic strength of the
lightcurves, as described above, we added all the bursts in three
groups; bursts 1-8; 9-16; and 17-22. We phase-align and stack the
bursts within each group, and therefore effectively analyze three
characteristic sets of pulse profiles. We used five energy bands for
each group.  The energy ranges are given in \S 2 above.

We compare our models directly with the data.  To do this, we compute
a model lightcurve of intensity and spectrum versus pulse phase.  In
this initial step we consider only the lightcurve of the hot spot, not
any background emission.  We fold the model spectrum through the {\it
RXTE} response matrix to get counts per channel as a function of
rotational phase.  For each channel range, we then add a
phase-independent background (which is a free parameter) and normalize
it so that the total number of counts in that channel range, summed
over all phases, is the same as the number of counts in that channel
range in the observed lightcurve.  Our final step is to shift the entire
lightcurve in phase (by same amount for all the channel bands)
to maximize the match with the observed
lightcurve, as determined by a $\chi^2$ statistic.

The procedure of adding backgrounds and normalizing counts
independently in each of the channel ranges has two implications.
First, the independent backgrounds mean that we leave the unpulsed
spectrum unconstrained.  Second, the independent normalizations mean
that although we calculate the initial spectrum and lightcurve of the
hot spot assuming a $kT=2$~keV blackbody, renormalization allows the
effective spectrum to deviate from blackbody. Therefore, we preserve the rms
variability of a blackbody with, e.g., Doppler shifts,
but not the whole spectrum. Indeed, in reality we
do not expect the temperature to be constant throughout a hot spot;
for example, if the hot spot is related to a magnetic pole, it might
be hotter in the center than at the edges.  In addition, the pulse
profiles extracted from the observations incorporate data over a
period of several tens of seconds, during which the hot spot is likely
to cool significantly.  Thus, the time-averaged spectrum will not
truly be a blackbody, hence we let the spectrum be adjusted by
allowing independent normalizations.

The initially chosen hot spot temperature of 2 keV is justified for
the following reasons: (1) It is expected to be close to the average
blackbody temperature of the hotspot for most of the bursts (see
Kuulkers et al. 2003 for blackbody temperature variation during
burst evolution for several sources),
and (2) The likelihood distributions (see next two paragraphs for
description) for the parameters are similar, even for widely different
initially chosen hot spot temperatures.

Using the above procedure of adding backgrounds, normalizing, and
shifting phases, we find the best $\chi^2$ value for many different
combinations of parameters. To be systematic, we consider combinations
of parameters on a grid (for the values of each parameter, see
Table 1). We also use the Markov Chain Monte Carlo (MCMC) method to ensure
that the grid method does not miss any significantly low $\chi^2$ values.
The number of counts per phase bin in each channel range
is large enough that Gaussian statistics apply, hence the likelihood
is proportional to $\exp[-\chi^2/2]$. We apply the physical constraint
that the value of each parameter is same for all the channel ranges
for a given burst group (some parameters, for example, the size and the 
location of the hot spot may change from one burst group to another). 
Therefore, for each parameter combination,
the likelihood is proportional to $\exp[-\Sigma\chi^2/2]$, where the
sum is over all the channel ranges. 

Once the likelihoods are computed for each combination of parameters,
we produce likelihoods for each parameter individually via the process
of marginalization.  In this process, let the posterior probability
density over the full set of model parameters $\theta_1\ldots\theta_n$
be $p(\theta_1\ldots\theta_n)$.  If we are interested in the credible
region for a single parameter $\theta_k$, then we integrate, or
marginalize, this probability distribution over the ``nuisance"
parameters $\theta_1\ldots\theta_{k-1}$ and
$\theta_{k+1}\ldots\theta_n$:
\begin{equation}
q(\theta_k)=\int d\theta_1\ldots d\theta_{k-1}
d\theta_{k+1}\ldots d\theta_n p(\theta_1\ldots\theta_n)\; .
\end{equation}
For our purposes we assume that each combination of parameters has the
same a priori probability, hence the posterior probability is simply
proportional to the likelihood. For Gaussian distribution of likelihoods
of a parameter, this method is equivalent to the standard frequentist
procedure. However, for non-Gaussian distribution, this Bayesian method is 
more general.

We find that the posterior probability distributions are not Gaussian
for most of our parameters.  Therefore, although there is a unique
maximum likelihood value for any parameter, credible regions cannot be
quoted uniquely.  For example, one gets a different value if one
assumes a symmetric distribution than if one looks for the smallest
region containing a certain total probability.  We have adopted a
variant of the latter procedure.  Using linear interpolations of the
probability between grid points for a given parameter, we compute the
smallest region that includes the maximum likelihood point and
contains 90\% of the total probability.  For parameters expected to
remain unchanged between bursts (specifically, $R/M$ and the observer
inclination $i$, and also the beaming parameter $n$) 
we combine data from all bursts.  For the other
parameters, we calculate likelihoods separately for each of the three
burst groups.

\section{Results and Conclusions} \label{sec:4}

As discussed in \S~3, we perform comparisons using two EOS
models. Figure 6 shows the goodness of fit for a certain burst group
and channel range, and for a representative combination of parameter
values. The $\chi^2$ value for this example comes out to be $\sim
25.5$, for the number of degrees of freedom equal to 24.  This
goodness of fit suggests that the overall model framework is a
reasonable representation of the data. 

In Figure 7, we plot the likelihoods with the parameter $R/M$, for
two EOS models, using the data from all 22 bursts. While we can not
determine an upper limit of $R/M$, lower limits $(4.7$ for A18$+\delta v+$UIX
and $4.2$ for A18) of $90$\% confidence regions (given in Table 2)
can be computed. We do not consider $R/M$
values greater than $6.0$, as for these values, the stellar mass is too
small $(< 1.07 M_\odot$ for A18 and $< 1.29 M_\odot$ for A18$+\delta v+$UIX)
to be realistic for an accreting neutron star. A simple extrapolation
of the likelihood curves in Figure 7 shows that the probability of 
$R/M < 3.6$ is very small. However, for $R/M < 3.52$ and for Schwarzschild
spacetime, photons from a
single point on the stellar surface may reach the observer by multiple
paths. As a result, the corresponding lightcurves may be qualitatively
different from the ones we calculate. Therefore, we can say that one
interpretation of the data is that $R/M$ is greater than 4.2 with
$90$\% confidence, but we can not completely exclude the possibility
of $R/M < 3.52$.

Figure 8 displays the likelihood distribution with the observer's
inclination angle $i$. This figure shows that $i > 22^{\rm o}$ with very high probability.
Here, as well as for the likelihood calculation of other parameters, we focus
our fitting procedure on the range $20^{\rm o} - 50^{\rm o}$ for $i$.
This is because, with a smaller number of grid points for other parameters,
our fitting for inclination angle values in the range $5^{\rm o} - 90^{\rm o}$
shows that the likelihood values for $i < 20^{\rm o}$ are very small, and
the inclusion of inclination angles greater than $50^{\rm o}$ produces only
minor changes to our results. However,
constraints on inclination angle by observations in other wavelengths would
help us constrain other parameters more effectively. Presently, M. Krauss et al. 
(2004, in preparation) are working on this using optical data.

We keep the beaming parameter $n$ (defined in \S~3) fixed in all three
burst groups in our statistical procedure. This is because
the thermonuclear flash is expected to occur at the bottom
of a pile of accreted matter. Therefore, although the original
emission may be isotropic, what comes out through the layers of
accreted matter should be beamed. As the optical depth of these
layers is expected to be very high, we expect the law
of darkening (and hence the value of $n$) to be the same for
all the bursts. 

The likelihood distributions for $n$ for two EOS models are shown in Figure 9.
For both EOS models, the beaming parameter $n$ peaks at $n=0.75$. This 
figure shows that $n$ is well constrained on lower side, and the lower limit
of the $90$\% confidence regions (for both the EOS model) is $0.55$ (given 
in Table 2). The law of darkening for semi-infinite 
plane-parallel layers with a constant net flux and
Thomson scattering is given in Chandrasekhar (1960). 
In Figure 10, we compare this law of darkening $I(\psi)$
with our emission function $\cos^n(\psi)$. We find that,
for $n=0.55$ (where likelihood value is $\sim 15-20$\% of the peak
likelihood value), our
emission function matches with $I(\psi)$ quite well, except for
nearly tangential emission (for which the emitted flux is small). The
matching is also reasonably good for the most probable value $n=0.75$. 
Therefore, the data are consistent with Thomson scattering through
an optically thick layer (probably in the accumulated accreted
matter) of thermal electrons.

In Table 3, we show the 90\% confidence intervals for the
polar angle of the center of the hot spot $(\theta_{\rm c})$ 
for both the EOS models. Here we
give the results for the three burst groups separately, as the values
of this parameter may change from one group to another. 
The union of 90\% confidence intervals for all burst groups and EOS
models is $60^{\rm o}-139^{\rm o}$. 
Assuming that the hot spot is at the
magnetic pole, this implies an angle of $40^{\rm o}-90^{\rm o}$ between
the spin axis and the magnetic axis. 
The peak of the likelihood distribution for $\theta_{\rm c}$ always occurs
around the interval $90^{\rm o}-110^{\rm o}$ (see Figures 11, 12 \& 13),
hence the magnetic pole may be close to the rotational equator.

Table 4 displays the likelihood values for the angular radius $(\Delta\theta)$
of the hot spot for different burst groups and EOS models. For the
burst group $1-8$, $\Delta\theta$ is comparatively better constrained,
and a smaller spot $(\Delta\theta \sim 5^{\rm o}-25^{\rm o})$ is likely.
For the other two burst groups, comparatively larger spots are probable.

A direct comparison of the likelihoods of two EOS models shows that
the difference is not enough to constrain EOS models by this direct method.

The pulsar mass function and orbital
period for this source are $0.002016\,M_\odot$ and 4.27 hours
respectively (Markwardt, Strohmayer, \& Swank 2003). Hence,
a reasonable range of neutron star mass $(1.4\,M_\odot - 2.0\,M_\odot)$
and our $i > 20^{\rm o}$ inclination angle constraint imply a companion mass of
$0.17\,M_\odot - 0.72\,M_\odot$.
For this range of masses and for the observed orbital
period, the companion can be neither a helium main sequence star nor
a degenerate star (Bhattacharya \& van den Heuvel 1991). 
It is also too large to be a brown dwarf (as was predicted for
another accreting millisecond pulsar system SAX J1808.4-3658; Bildsten
\& Chakrabarty 2001). The most
probable option is that it is either a hydrogen burning main sequence star, or
an evolved (sub-)giant star. However, the companion mass is too small
for it to have evolved off the main sequence (B\"ohm-Vitense 1992).
Therefore, the most likely option is a hydrogen main sequence star.
This is also seen in Figure 14. If the companion is a normal hydrogen 
burning main sequence star (as shown in Figure 14), its maximum possible
mass is $\sim 0.48 M_\odot$. This is because, the radius of the
companion can not be smaller than that of a hydrogen main sequence star (for
a given mass), but it can be bigger, because it can be bloated due to X-ray
heating by the neutron star and/or the accretion disk. This upper limit of
companion mass corresponds to a lower limit of observer's inclination angle
$24^{\rm o}$ (using the pulsar mass function value and a lower limit
of neutron star mass $1.4 M_\odot$). Interestingly, this lower limit of 
inclination angle is close to that found by our lightcurve fitting.
However, if the mass fraction of hydrogen in the hydrogen burning main 
sequence companion star is very low, then it may have a higher mass than
$0.48 M_\odot$ (see Rappaport \& Joss 1984).
If we consider the lower limit $(0.17\,M_\odot)$ of the companion mass,
then its radius is about $98$\% bigger than the normal radius 
(Bhattacharya \& van den Heuvel 1991).

In conclusion, the significant overtones observed in the pulse
profiles of burst brightness oscillations from XTE~J1814--338 open up
new possibilities for constraints on neutron star structure as well as
on source geometry and emission properties. 
In a subsequent paper, we will study the analysis possible with future
large-area detectors, in particular whether EOS models may be constrained
strongly based on their likelihoods.

\acknowledgements

We thank Arun V. Thampan for providing the rapidly spinning neutron
star structure calculation code, and Geoff Ravenhall for supplying the
tabulations of nucleonic equations of state. 
We also thank Deepto Chakrabarty and Miriam Krauss for notifying us about 
their unpublished work.
We acknowledge John Miller, Didier 
Barret, and Biswajit Paul for reading the paper and sending their comments.
Many of the results in this paper were obtained using the
Beowulf cluster of the department of astronomy, University of Maryland 
(courtesy of Derek C. Richardson). This work was supported in part by
NSF grant AST~0098436.

{}

\section*{Appendix}

We use three codes in series to calculate the $\chi^2$ values.
These codes have been checked in different ways. Here we mention 
a few of them. The ray-tracing
code traces back the paths of photons from the observer to the 
surface of the star, and calculates different parameter values 
at the footprints of the photons on the
surface. For a photon that is 
emitted from the surface tangentially (i.e., the one with maximum 
angular momentum), one can calculate the amount of
angular deflection at infinity (for the Schwarzschild spacetime). 
The comparison between these
values with the outputs from our code give satisfactory
results. For example, when $R/M = 4.2$, the total angle traveled by a 
photon emitted tangentially at the surface is $145^{\rm o}.8$, and our
code yields $145^{\rm o}.1$.
The code also calculates the total redshift with typically less
than $0.01$\% error. However, a photon travelling 
directly over a pole of the star gives a few degrees of error in the
$\theta$-position of its emission point. But the error in this single ray
does not introduce significant error in our final results.

For both the Schwarzschild and Kerr metrics, the value of 
$L = -p_{\phi}/p_t$ can be derived analytically
at the observer's position and at the surface of the star. 
Here $p_{\phi}$ and $-p_t$ are photon's
$\phi$-angular momentum and energy respectively. At the observer's
position
\begin{equation}
L = -b\sin i \sin \alpha \; ,
\end{equation}
where, $b$ and $\alpha$ are plane polar coordinates in observer's sky,
and $i$ is the observer's inclination angle. At the surface of the star,
\begin{equation}
L = \frac{g_{\phi\phi}^{1/2} \cos\xi}{\sqrt{\frac{g_{\phi t}^2}{4 g_{\phi\phi}}-g_{tt}}-\frac{g_{\phi t}}{2 g_{\phi\phi}^{1/2}} \cos\xi} \; ,
\end{equation}
where, $\xi$ is the emission angle with $\phi$-direction, and $g_{\mu\nu}$'s
are the metric coefficients of a metric
\begin{equation}
dS^2 = g_{tt} dt^2 + g_{rr} dr^2 + g_{\theta\theta} d\theta^2 + g_{\phi\phi} d\phi^2 + g_{\phi t} d\phi dt \; .
\end{equation}
As $L$ remains constant along the photon's path, the values of $L$ from
equations (2) and (3) should be the same. We find that with our ray-tracing 
code the maximum difference in these two values is $\sim 1$\%.

The lightcurve calculation code computes the energy dependent observed
flux from different hotspots on the stellar surface. This is done
by the integration of observed specific intensity over the image of the
hotspot at the observer's sky. As the image of the hotspot is of 
irregular shape, this integration is done using a Monte Carlo method. We
ensure that different random seed values and different number of
points (used for Monte Carlo integration) give a stable result. 
For example, for parameter combinations giving $\chi^2$/dof $\sim 1$, 
differences in $\chi^2$s are less than
$0.5$\%. For larger $\chi^2$ values, these differences are more, but
still very low, and these have little effect because it is the lower 
$\chi^2$ values that dominate the likelihoods. These differences 
in $\chi^2$s appear to be more or less random, hence the differences
in total $\chi^2$s (addition for different channel ranges and
burst groups) are also small.
The integrated area is also verified to be correct (with typically less than
$0.1$\% error) even for a small spot (for example, with $5^{\rm o}$ angular 
radius) with its center at the same $\phi$ and $\theta$ values
as of the observer.

Finally, the comparison code compares the theoretical lightcurves with the
observed data, and computes the $\chi^2$ values. We check
this code in different ways. For example, we consider different
maximum background values and different steps in background values
to ensure a stable result (maximum $\sim 0.03$\% difference in $\chi^2$
values). We also use synthetic data to confirm that the known parameters
are obtained to accuracies compatible with our uncertainty region for the
real data.

\clearpage
\begin{figure}
\epsscale{1.0}
\plotone{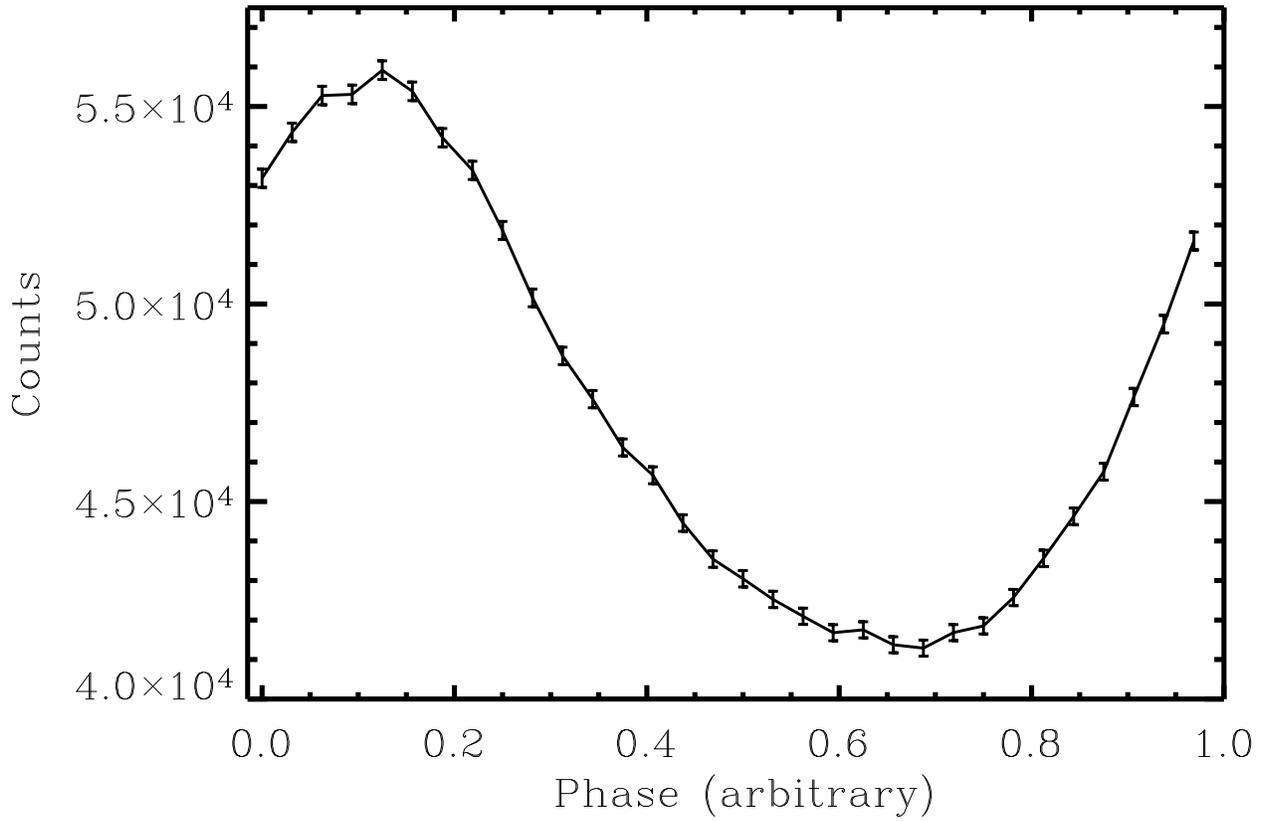}
\caption{The fully added observational lightcurve (22 bursts and
$0 - 28$ channels).\label{fig1}}
\end{figure}

\clearpage
\begin{figure}
\epsscale{1.0}
\plotone{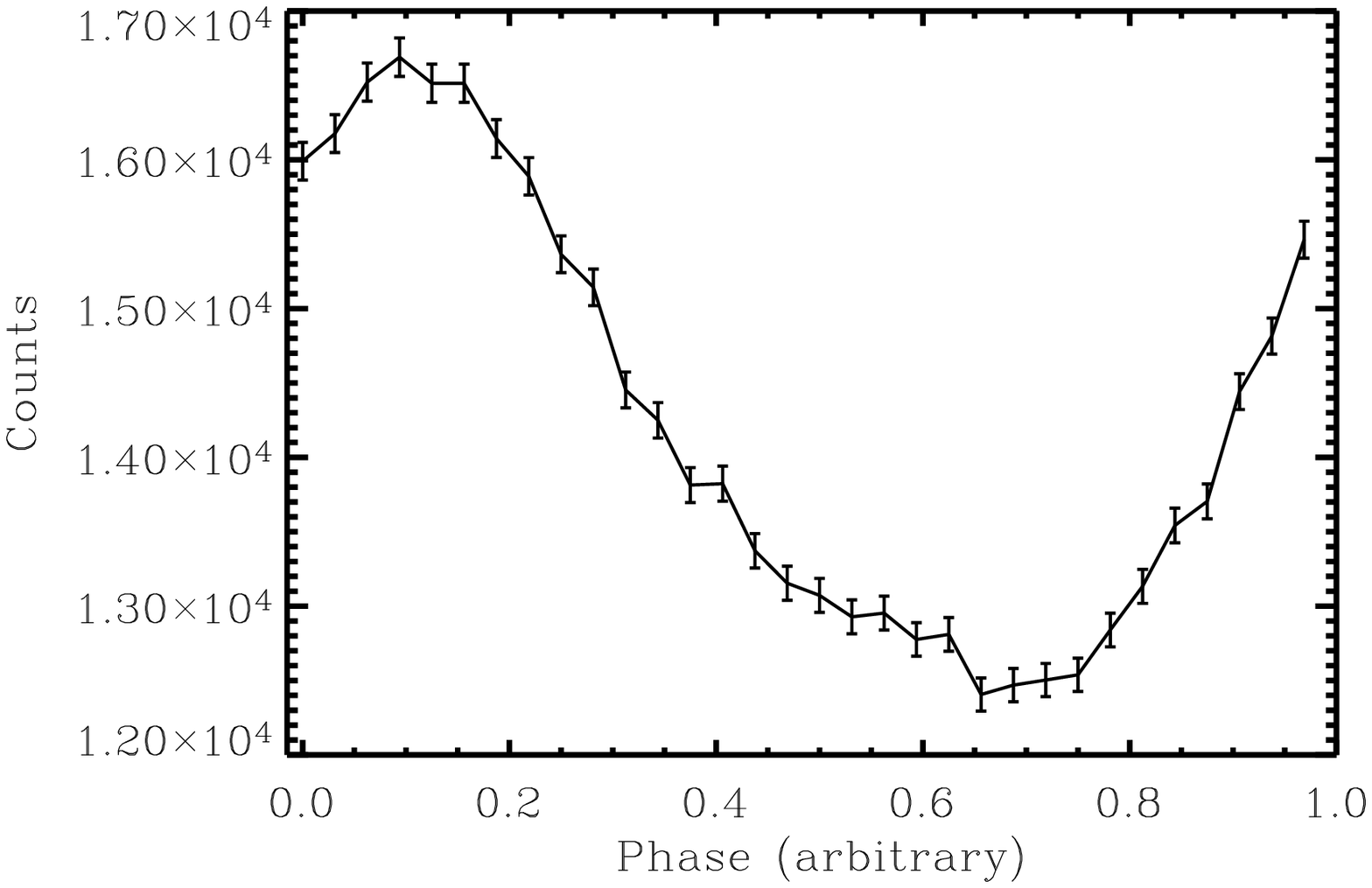}
\caption{The observational lightcurve (bursts: $1-8$ and
$0 - 28$ channels).\label{fig2}}
\end{figure}

\clearpage
\begin{figure}
\epsscale{1.0}
\plotone{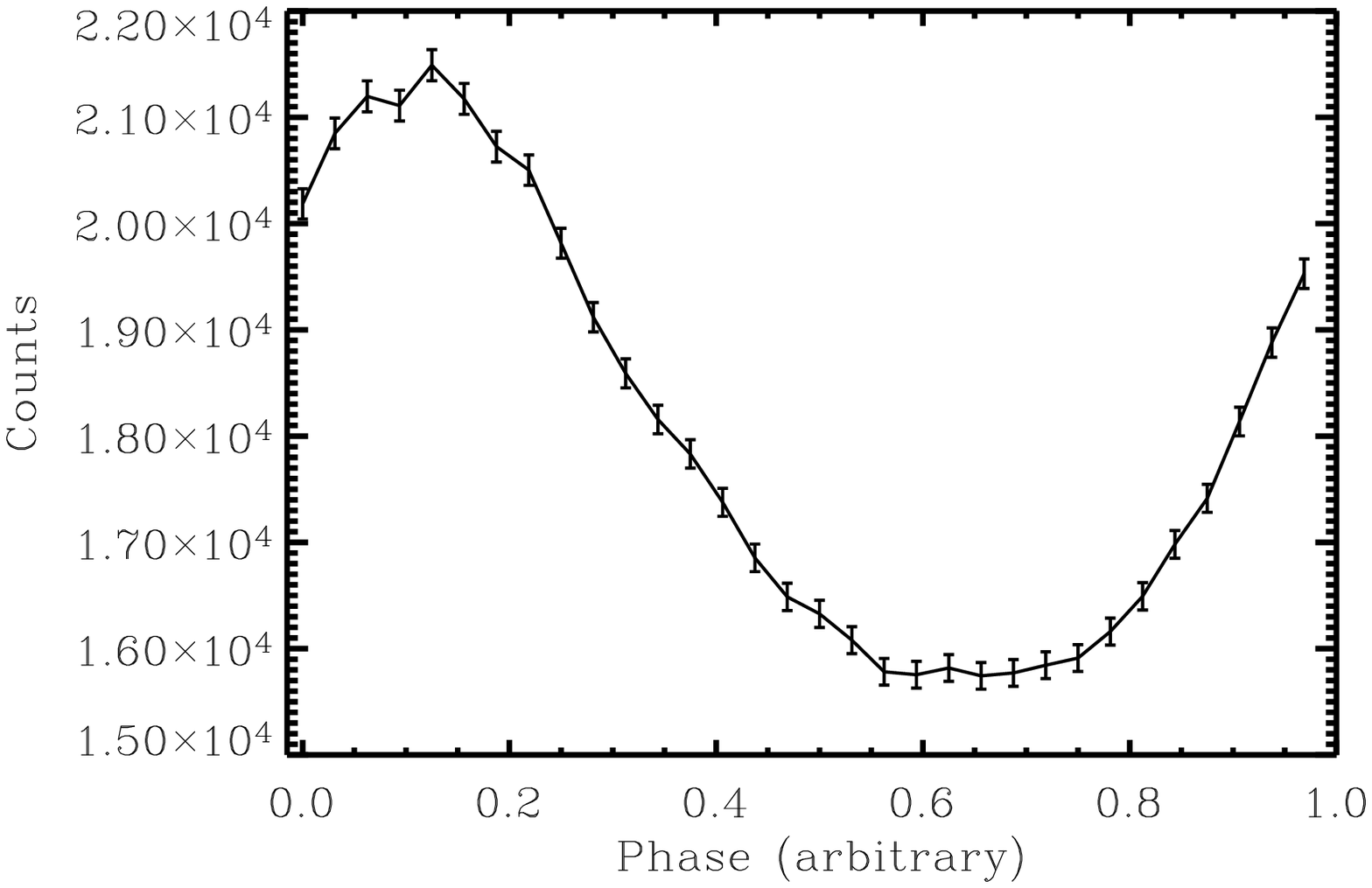}
\caption{The observational lightcurve (bursts: $9-16$ and
$0 - 28$ channels).\label{fig3}}
\end{figure}

\clearpage
\begin{figure}
\epsscale{1.0}
\plotone{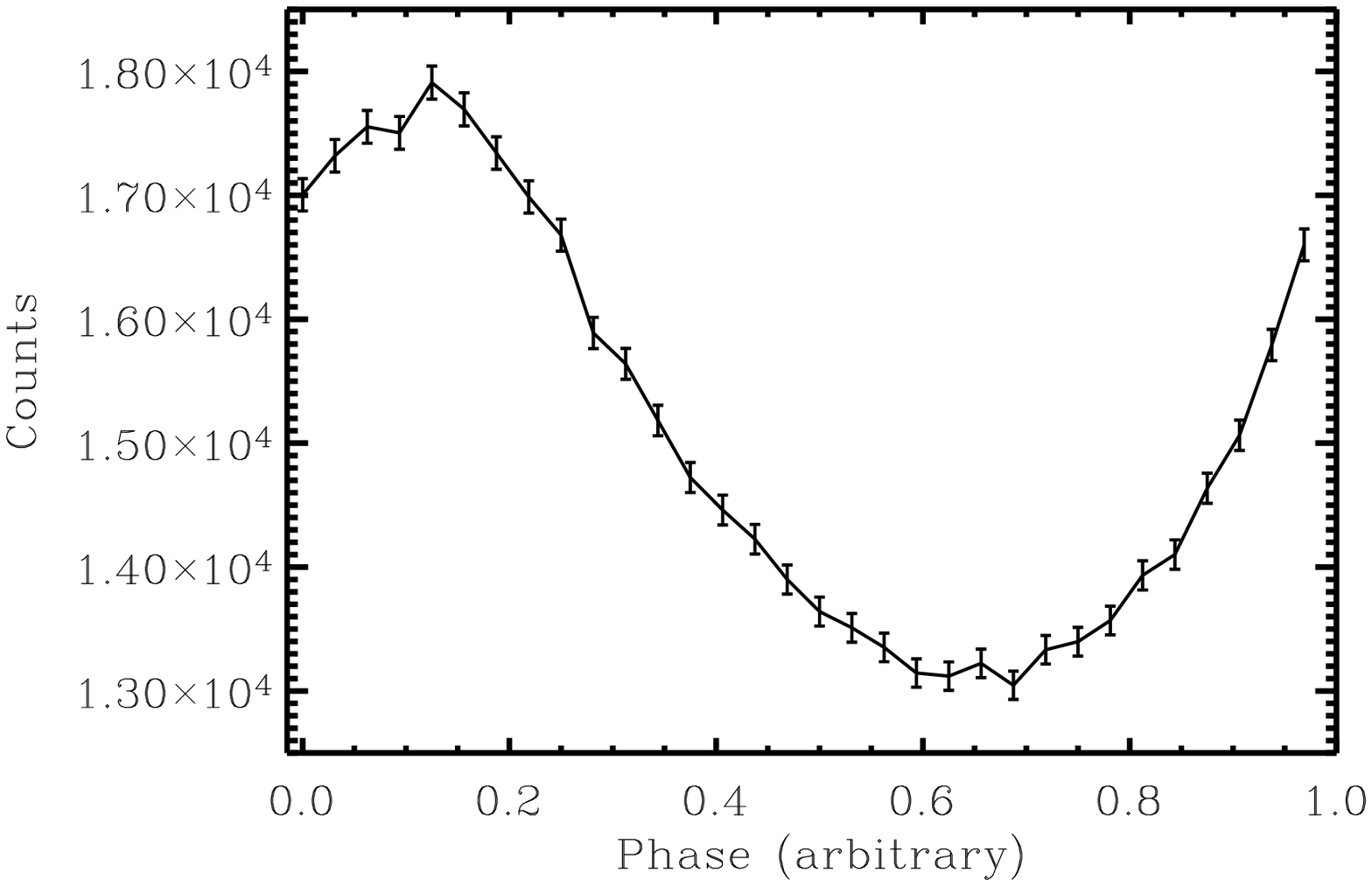}
\caption{The observational lightcurve (bursts: $17-22$ and
$0 - 28$ channels).\label{fig4}}
\end{figure}

\clearpage
\begin{figure}
\epsscale{1.0}
\plotone{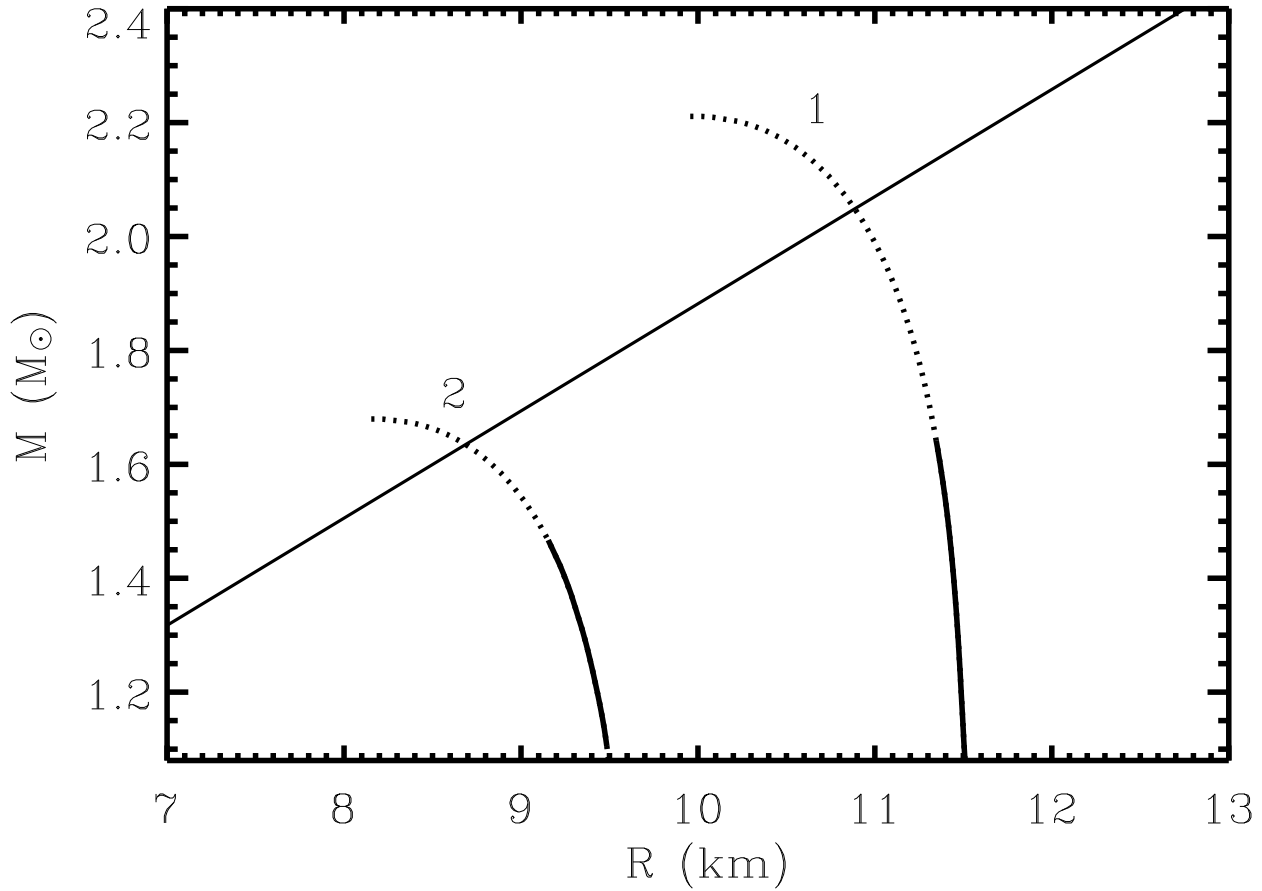}
\caption{Mass vs. radius diagrams: Curve 1 is for the EOS
model A18$+\delta v+$UIX and curve 2 is for
the EOS model A18. These diagrams are for stable
stellar configurations, spinning with a frequency $\sim 314$~Hz.
The solid parts of the curves (correspond to higher $R/M$ sides of the 
vertical lines in Figure 7) are the allowed regions with
90\% confidence interval of $R/M$. These regions
constrain the mass vs. radius plane for the neutron star in
XTE J1814-338. The solid oblique line corresponds to $R/M = 3.6$. In this
work, our results are for the $M-R$ space below this line.\label{fig5}}
\end{figure}

\clearpage
\begin{figure}
\epsscale{1.0}
\plotone{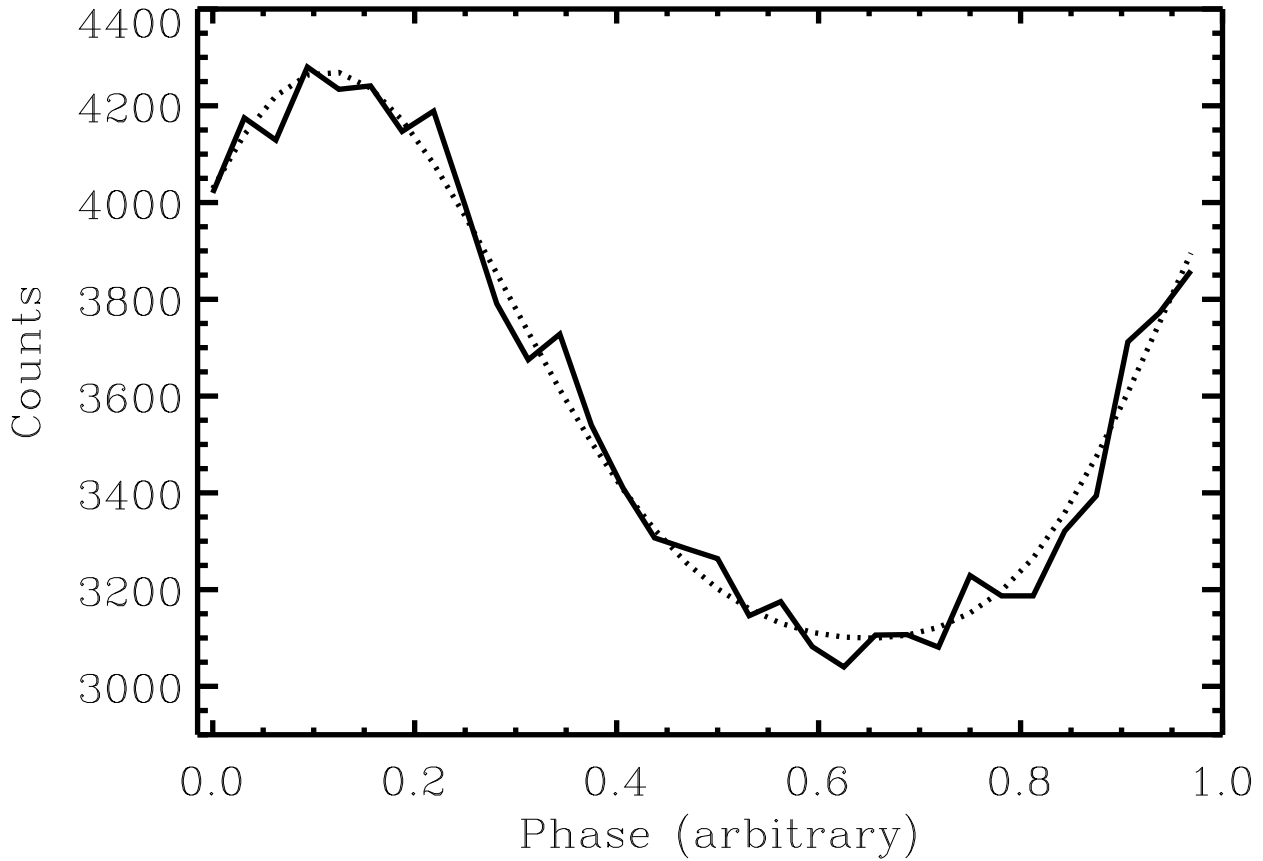}
\caption{An example fit to the data. Stacked data (solid
curve) for bursts $9-16$ (channel range: $7-10$) from XTE J1814-338,
versus a model (dotted curve) in which $R/M = 4.9$, $i=36^{\rm o}$,
$\theta_{\rm c}=110^{\rm o}$, the hot spot has an angular radius of
$45^{\rm o}$, and $n=0.8$ (see text for
description of parameters). Here we have used the EOS model
A18$+\delta v+$UIX. In this fitting we get a $\chi^2$ value $\sim
25.5$, for the number of degrees of freedom equal to 24.\label{fig6}}
\end{figure}

\clearpage
\begin{figure}
\epsscale{1.0}
\plotone{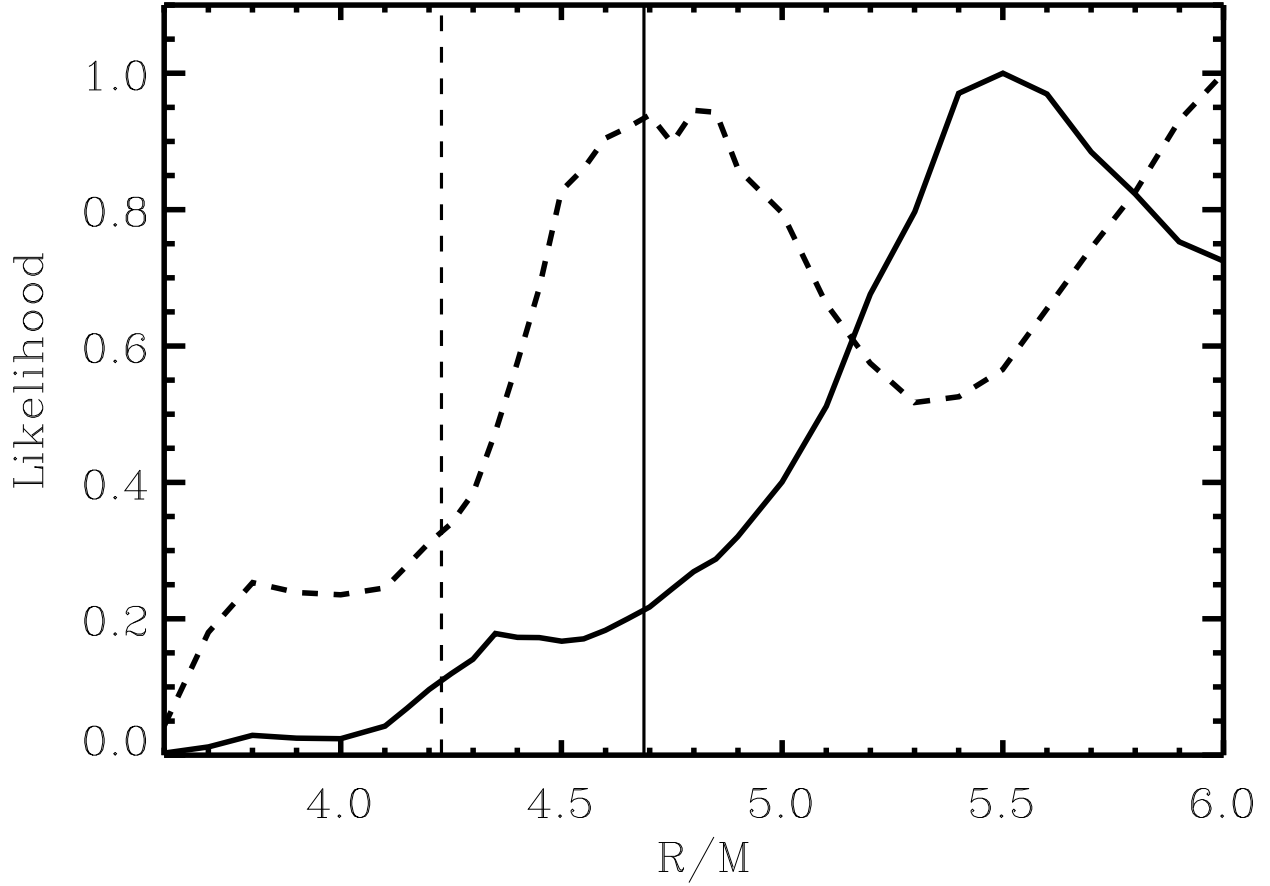}
\caption{Likelihood distribution of stellar radius to mass ratio
$R/M$, using data from all 22 bursts.  The solid curve is for the EOS
model A18$+\delta v+$UIX, and the dashed curve is for
the EOS model A18. The solid vertical line gives the lower limit of
the 90\% confidence region for the EOS model A18$+\delta v+$UIX,
and the dashed vertical line gives that for A18.\label{fig7}}
\end{figure}

\clearpage
\begin{figure}
\epsscale{1.0}
\plotone{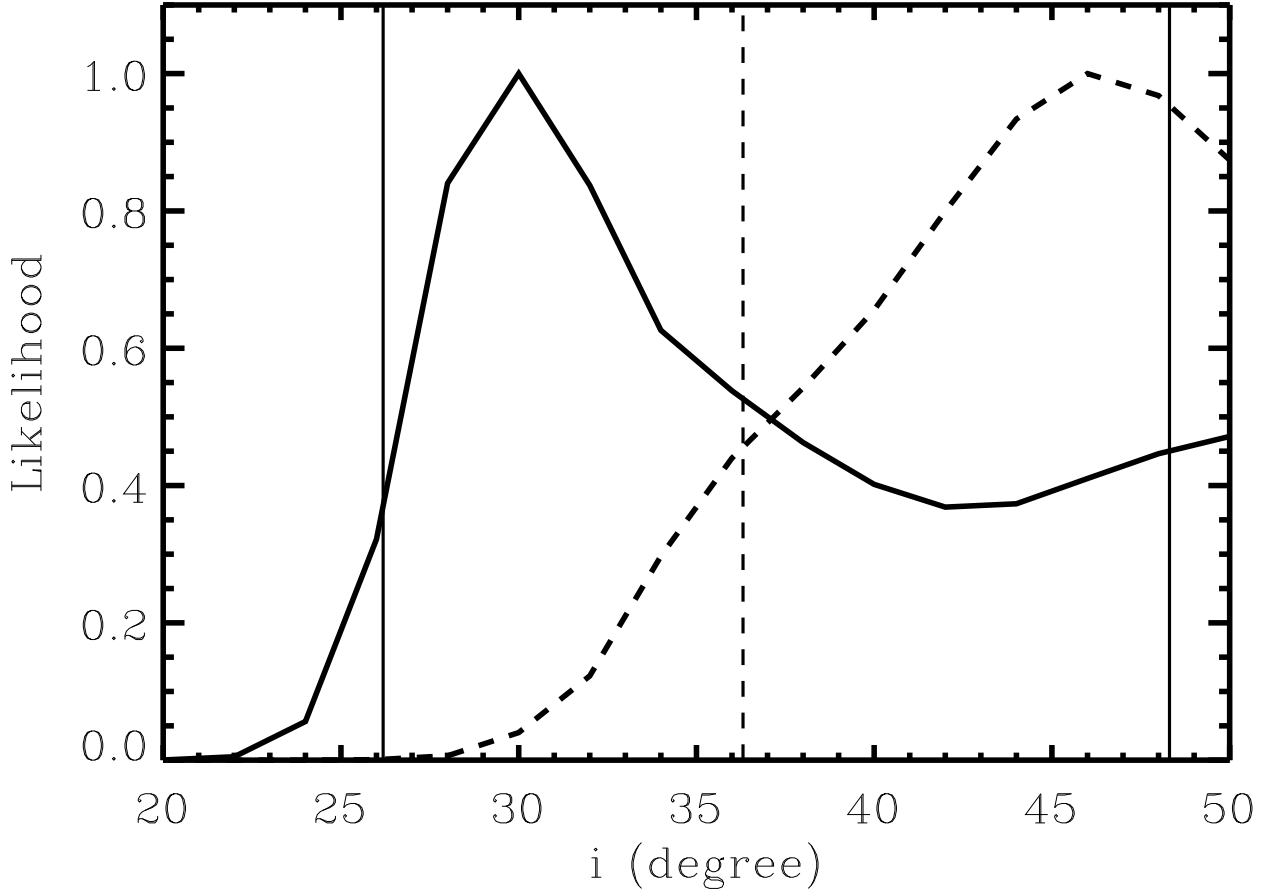}
\caption{Likelihood distribution of observer's inclination angle $i$,
using data from all 22 bursts. Solid curve is for the EOS model
A18$+\delta v+$UIX, and the dashed curve is for the EOS
model A18. The solid vertical lines give
90\% confidence interval for the EOS model A18$+\delta v+$UIX, while
the dashed  vertical line gives
the lower limit of 90\% confidence region for A18.
This figure demonstrates that a value of $i$ less
than $22^{\rm o}$ is highly improbable.\label{fig8}}
\end{figure}

\clearpage
\begin{figure}
\epsscale{1.0}
\plotone{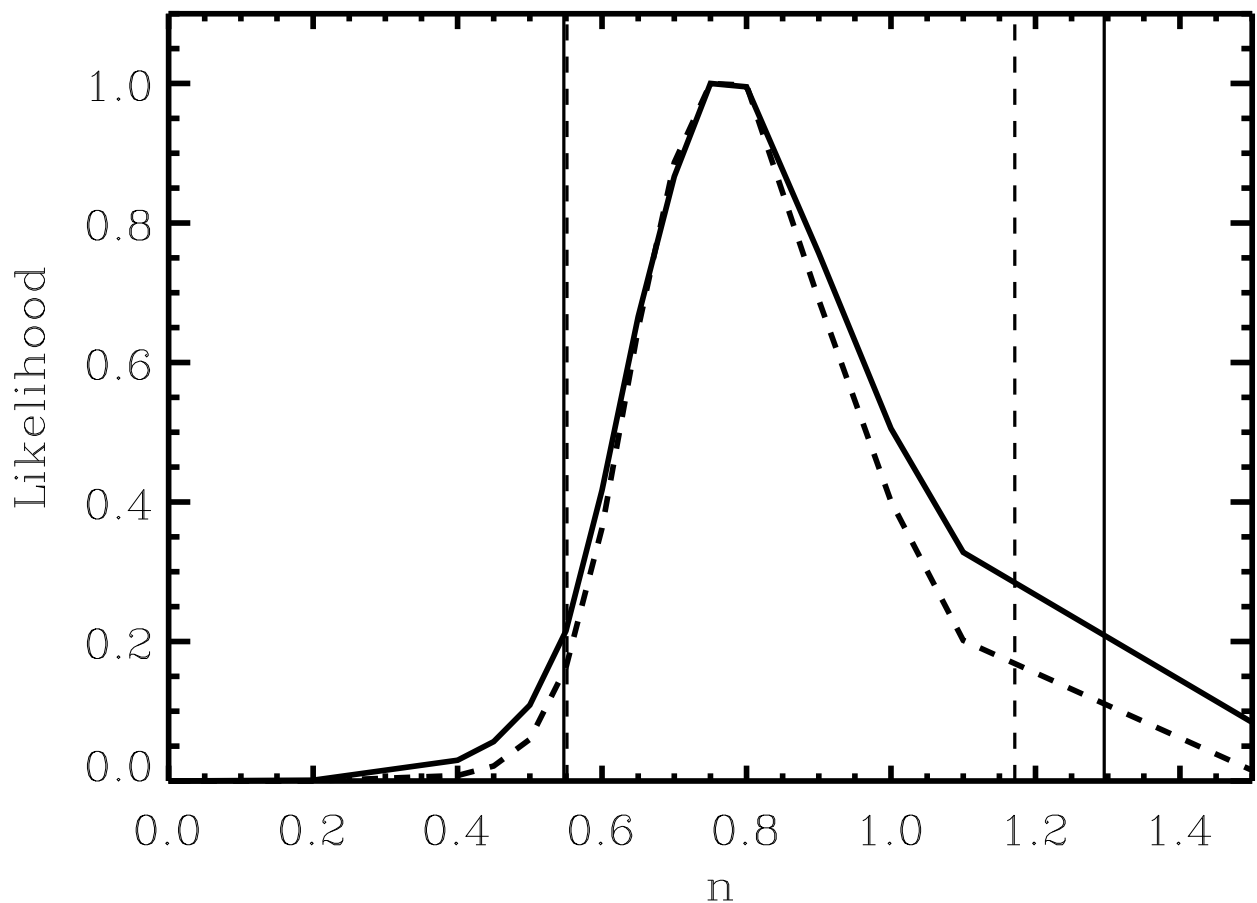}
\caption{Likelihood distribution of the beaming parameter $n$ (defined
in \S~3), using data from all 22 bursts. The solid curve is for the EOS
model A18$+\delta v+$UIX, and the dashed curve is for
the EOS model A18. The solid vertical lines give 90\% confidence interval
for the EOS model A18$+\delta v+$UIX, while the dashed  vertical
lines give that for A18. This figure demonstrates that a value of $n$ less 
than $0.2$ is highly improbable.\label{fig9}}
\end{figure}

\clearpage
\begin{figure}
\epsscale{1.0}
\plotone{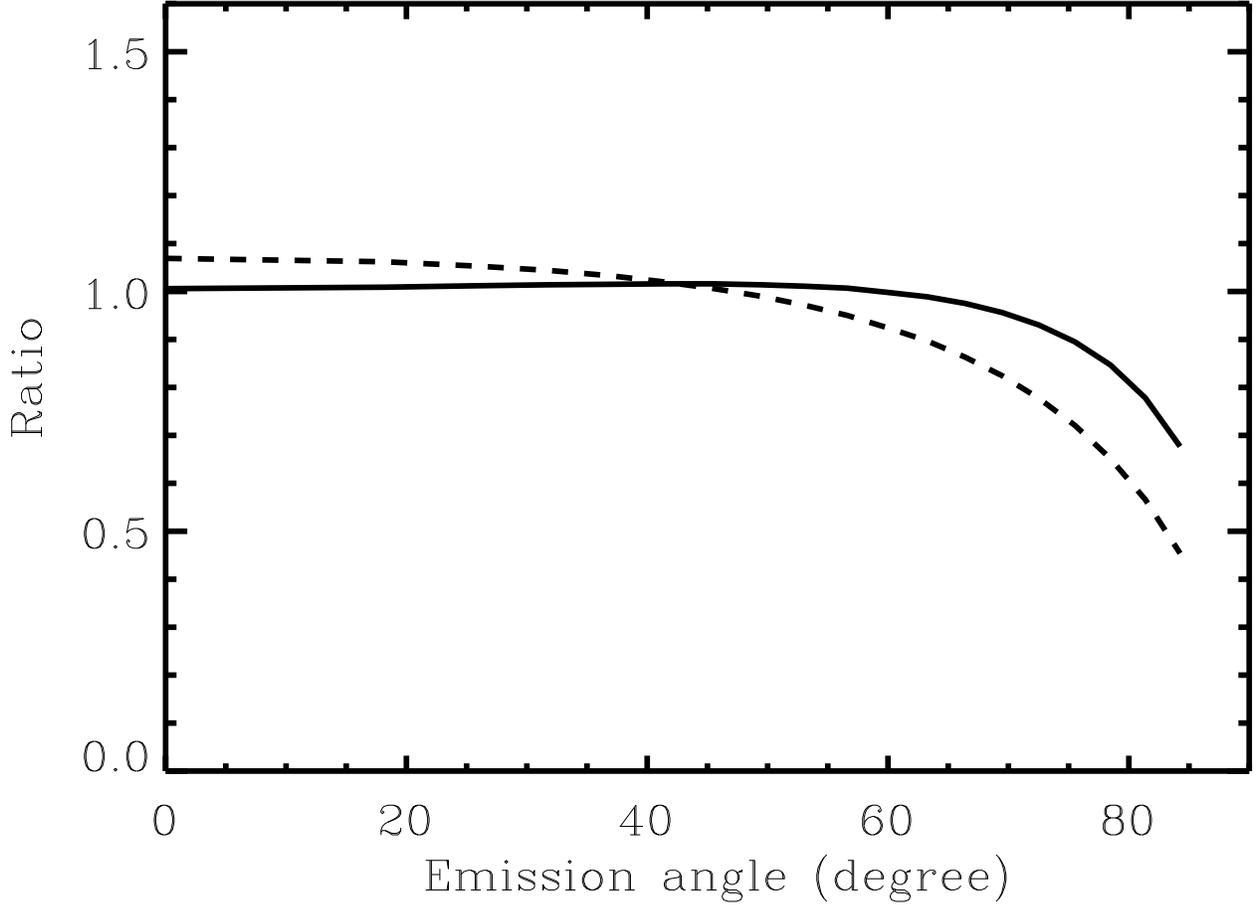}
\caption{Comparison between $I(\psi)$ (Law of darkening; Chandrasekhar 1960)
and our emission function $(\cos^{n}\psi)$: Both the functions give
the angular distribution of specific intensity in the emitter's
frame (corotating with the star).
Here the independent parameter is the emission angle $(\psi)$ 
(measured from surface normal) in the emitter's frame, and the dependent
parameter is the ratio of ${\mbox A}\cos^n(\psi)$ to $I(\psi)$
(where, A is an arbitrary constant, and
$I(\psi)$ is the law of darkening
for the radiative transfer in a semi-infinite plane-parallel
atmosphere with a constant net flux and Thomson scattering).
The solid curve is for $n=0.55$, and the dashed curve is for $n=0.75$. For
the former one, $I(\psi)$ is well fitted by our emission function,
except for very high emission angle, for which the emitted flux is
small.\label{fig10}}
\end{figure}

\clearpage
\begin{figure}
\epsscale{1.0}
\plotone{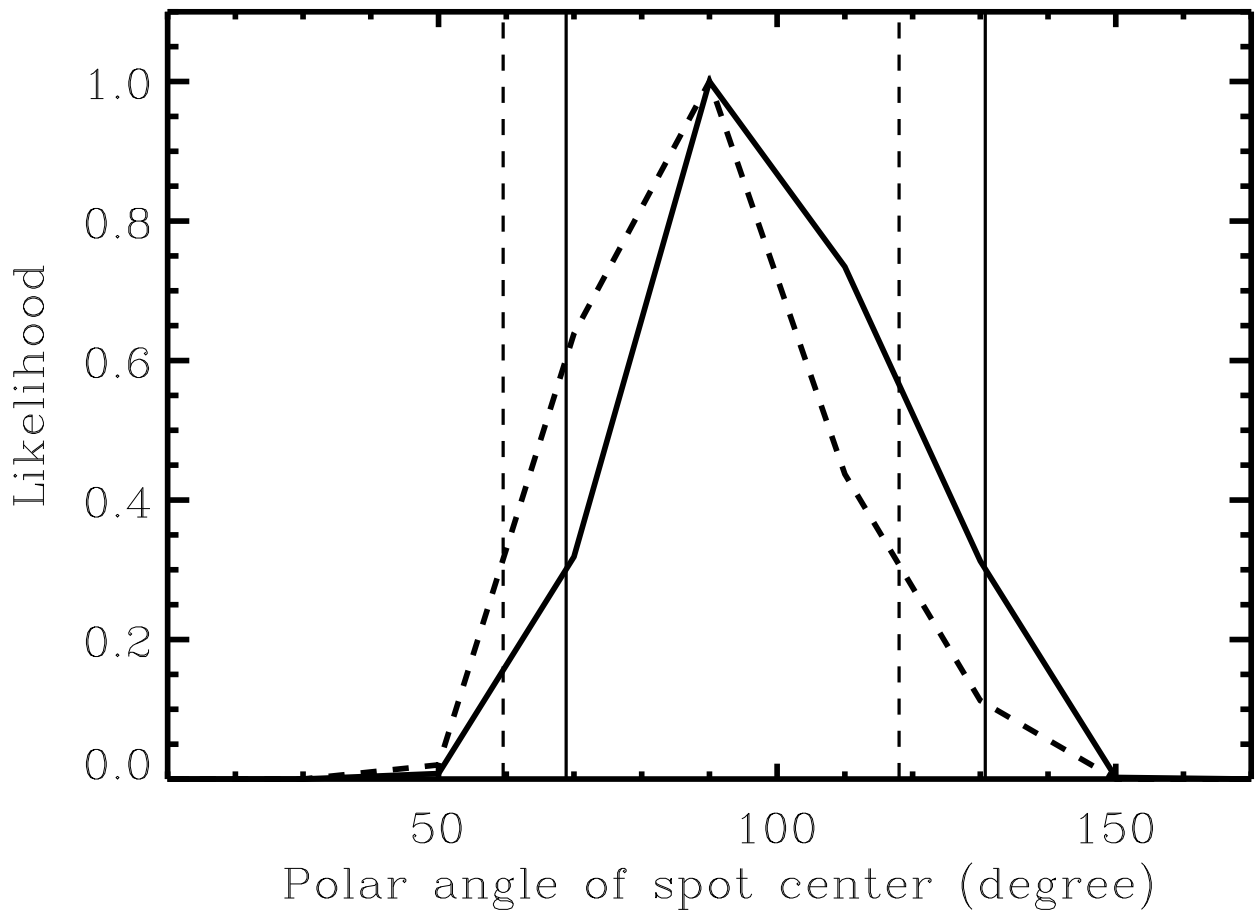}
\caption{Likelihood distribution of the polar angle of the center
of the spot $(\theta_c)$ for bursts $1-8$. Solid curve is for the EOS model
A18$+\delta v+$UIX, and the dashed curve is for the EOS model A18.
The solid vertical lines give 90\% confidence interval for the EOS model 
A18$+\delta v+$UIX, while the dashed  vertical lines give that for A18.
This figure demonstrates that a value of $\theta_c$ less than $50^{\rm o}$
or greater than $150^{\rm o}$ is highly improbable for bursts $1-8$.
\label{fig11}}
\end{figure}

\clearpage
\begin{figure}
\epsscale{1.0}
\plotone{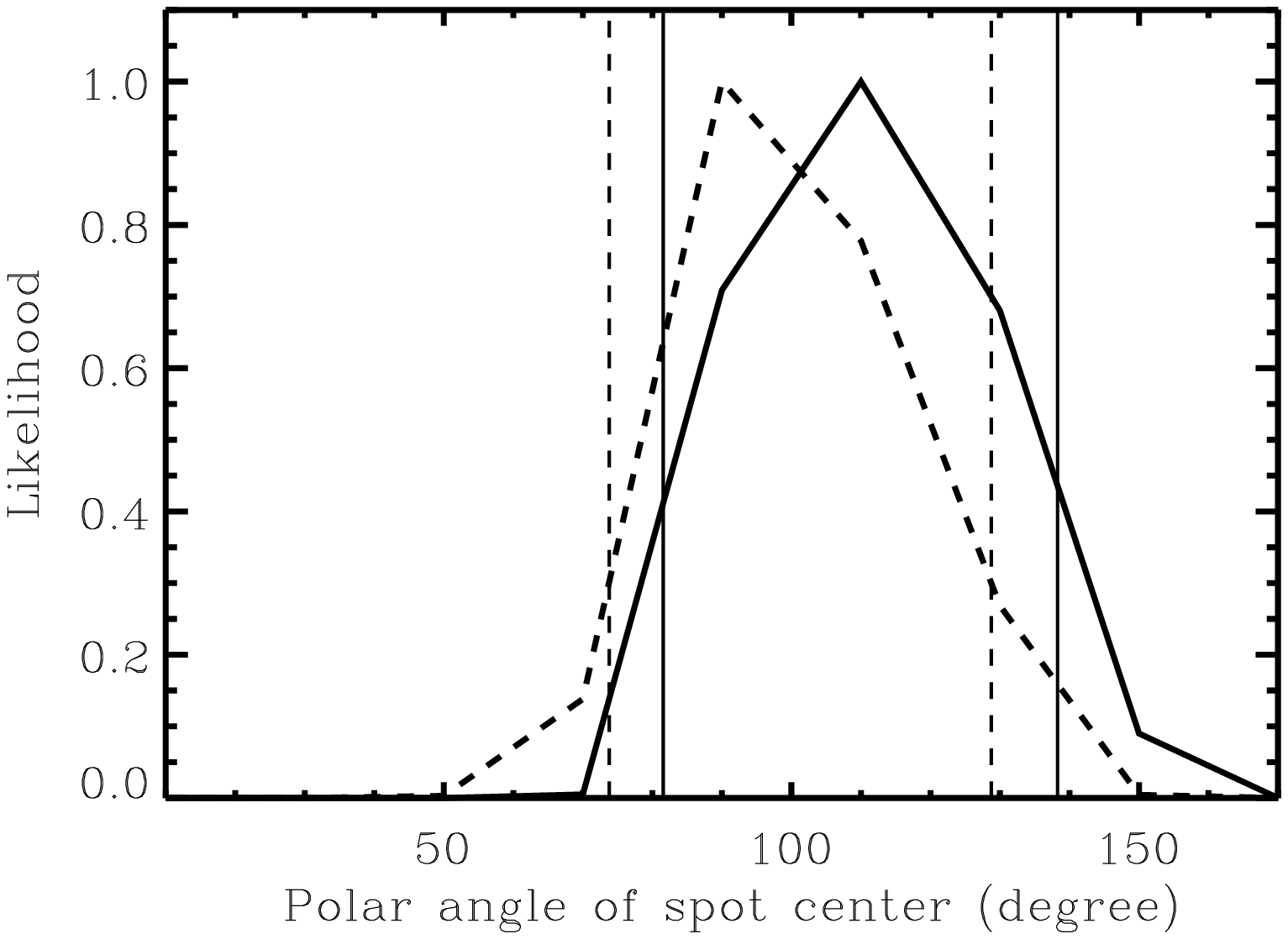}
\caption{Likelihood distribution of the polar angle of the center
of the spot $(\theta_c)$ for bursts $9-16$. Meanings of all the curves and 
lines are same as in Figure 11. This figure demonstrates that a value of 
$\theta_c$ less than $50^{\rm o}$ is highly improbable for bursts $9-16$.
\label{fig12}}
\end{figure} 

\clearpage
\begin{figure}
\epsscale{1.0}
\plotone{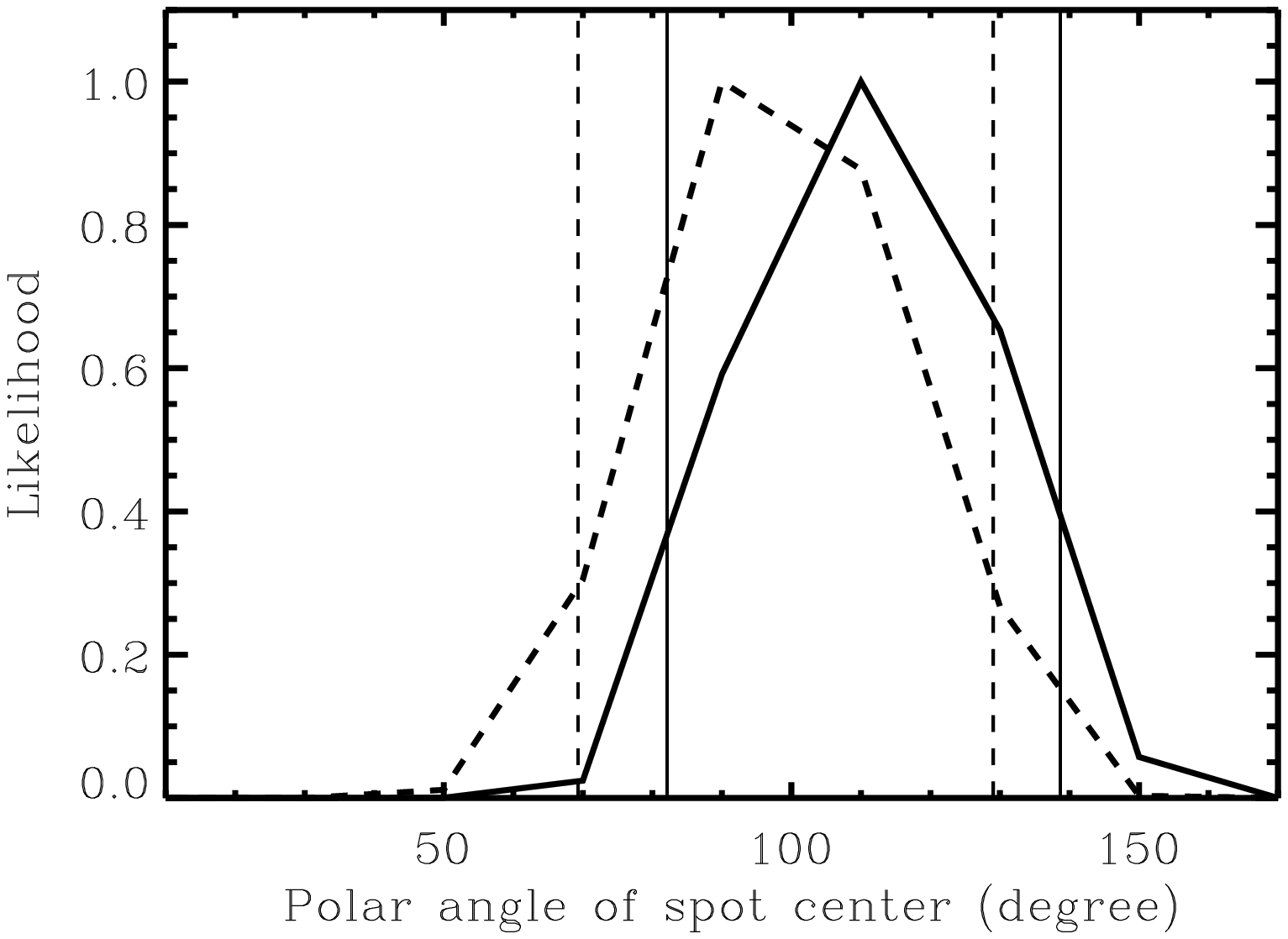}
\caption{Likelihood distribution of the polar angle of the center
of the spot $(\theta_c)$ for bursts $17-22$. Meanings of all the curves and 
lines are same as in Figure 11. This figure demonstrates that a value of
$\theta_c$ less than $50^{\rm o}$ is highly improbable for bursts $17-22$.
\label{fig13}}
\end{figure} 

\clearpage
\begin{figure}
\epsscale{1.0}
\plotone{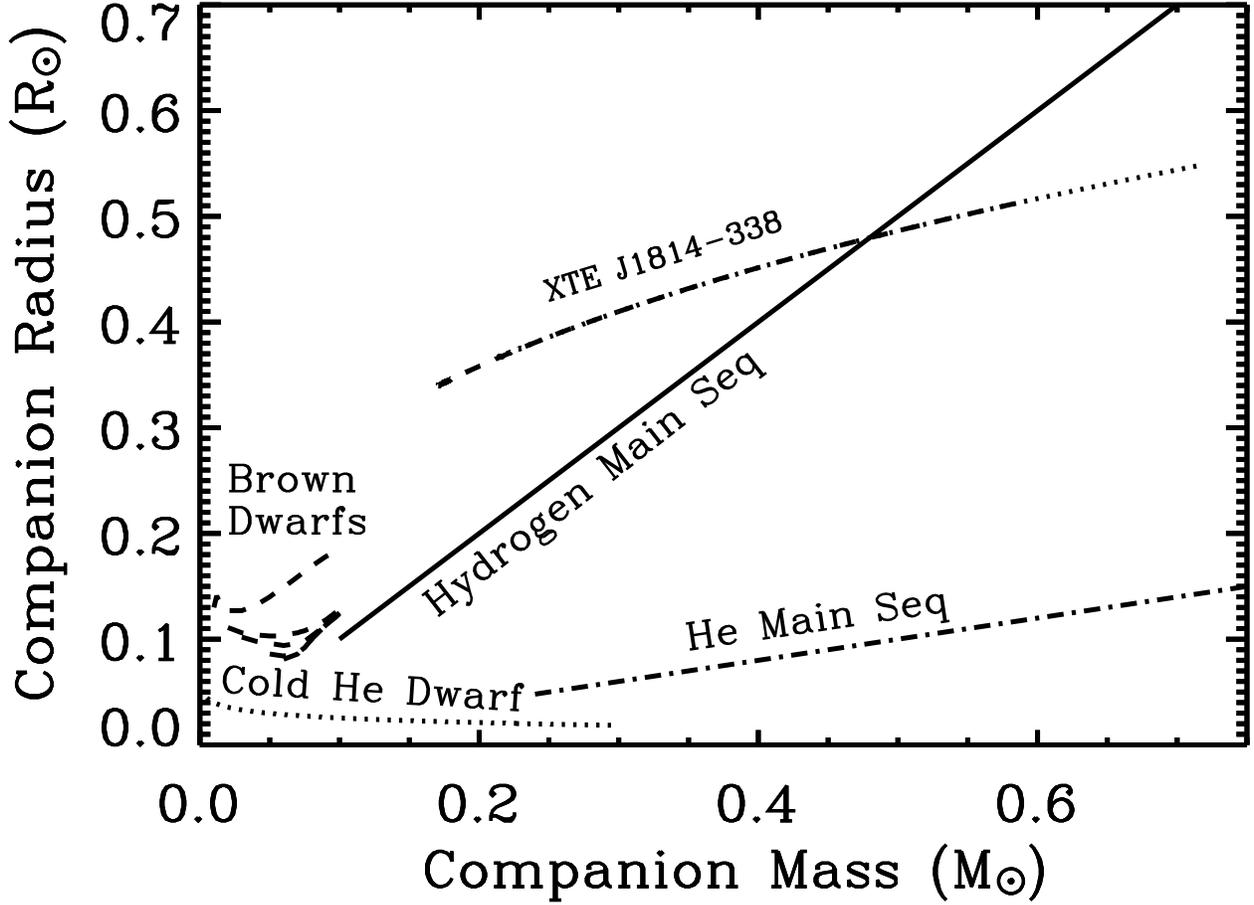}
\caption{Companion mass vs. radius plane (similar to the Figure 4 of
Markwardt et al. 2002): The lines under
the phrase `XTE J1814-338' are for the companion star in the source
XTE J1814-338 (for an observer inclination angle
$> 20^{\rm o})$; the dashed line is for a neutron
star mass $1.4 M_\odot$ and the dotted line is for the neutron
star mass $2.0 M_\odot$. For the purpose of comparison, we plot the
mass-radius curves for hydrogen burning main sequence star (solid line; 
using $R/R_{\odot} = M/M_{\odot}$),
helium main sequence star (dash-dot line; Verbunt 1990),
brown dwarfs (dashed lines), and cold helium dwarf (dotted line).
Brown dwarf models
are for ages of 0.1, 0.5, 1, 5, and 10 billion years ({\it from top to
bottom}). This figure demonstrates that the companion star in the source
XTE J1814-338 is likely to be a hydrogen burning main sequence star.\label{fig14}}
\end{figure}

\clearpage

\begin{deluxetable}{cl}
\tablecolumns{2}
\tablewidth{0pc}
\tablecaption{Grid values considered for the parameters for likelihood calculations.}
\tablehead{Parameter & Grid values}
\startdata
 & 3.60, 3.70, 3.80, 3.90, 4.00, 4.10, \\
 & 4.15, 4.20, 4.25, 4.30, 4.35, 4.40, \\ 
$R/M$ 	& 4.45, 4.50, 4.55, 4.60, 4.65, 4.70, \\
	& 4.75, 4.80, 4.85, 4.90, 5.00, 5.10, \\
	& 5.20, 5.30, 5.40, 5.50, 5.60, 5.70, \\
	& 5.80, 5.90, 6.00 \\
\hline
 & $20^{\rm o}$, $22^{\rm o}$, $24^{\rm o}$, $26^{\rm o}$, $28^{\rm o}$, $30^{\rm o}$, $32^{\rm o}$, \\
$i$  & $34^{\rm o}$, $36^{\rm o}$, $38^{\rm o}$, $40^{\rm o}$, $42^{\rm o}$, $44^{\rm o}$, $46^{\rm o}$, \\
 & $48^{\rm o}$, $50^{\rm o}$ \\
\hline
$\theta_{\rm c}$ & $10^{\rm o}$, $30^{\rm o}$, $50^{\rm o}$, $70^{\rm o}$, $90^{\rm o}$, \\
 & $110^{\rm o}$, $130^{\rm o}$, $150^{\rm o}$, $170^{\rm o}$ \\
\hline
$\Delta\theta$ & $5^{\rm o}$, $25^{\rm o}$, $45^{\rm o}$, $60^{\rm o}$ \\
\hline
 & $0.00$, $0.20$, $0.40$, $0.45$, $0.50$, \\
$n$  & $0.55$, $0.60$, $0.65$, $0.70$, $0.75$, \\
 & $0.80$, $0.90$, $1.00$, $1.10$, $1.50$ \\
\enddata
\end{deluxetable}

\clearpage

\begin{deluxetable}{cccc}
\tablecolumns{4}
\tablewidth{0pc}
\tablecaption{90\% confidence intervals for $R/M$, $i$ and $n$.}
\tablehead{Parameter & EOS & Lower limit & Upper limit}
\startdata
$R/M$ & A18$+\delta v+$UIX & 4.7 & 6.0 \\ \cline{2-4}
  &  A18 & 4.2 & 6.0 \\
\hline
$i$  & A18$+\delta v+$UIX & $26^{\rm o}$ & $48^{\rm o}$ \\ \cline{2-4}
  &  A18 & $36^{\rm o}$ & $50^{\rm o}$ \\
\hline
$n$  & A18$+\delta v+$UIX & 0.55 & 1.30 \\ \cline{2-4}
  &  A18 & 0.55 & 1.17 \\
\enddata
\end{deluxetable}

\clearpage

\begin{deluxetable}{cclcc}
\tablecolumns{5}
\tablewidth{0pc}
\tablecaption{90\% confidence intervals for the polar angle $(\theta_c)$ of the center of the spot.}
\tablehead{Parameter & EOS & Burst group & Lower limit\tablenotemark{a} & Upper limit\tablenotemark{a}}
\startdata
 & & Bursts: $1-8$ & $69^{\rm o}$ & $131^{\rm o}$ \\ \cline{3-5}
 & A18$+\delta v+$UIX & Bursts: $9-16$ & $82^{\rm o}$ & $138^{\rm o}$ \\ 
 \cline{3-5}
$\theta_{\rm c}$ & & Bursts: $17-22$ & $82^{\rm o}$ & $139^{\rm o}$ \\ \cline{2-5}
 & & Bursts: $1-8$ & $60^{\rm o}$ & $118^{\rm o}$ \\ \cline{3-5}
 & A18 & Bursts: $9-16$ & $74^{\rm o}$ & $129^{\rm o}$ \\ \cline{3-5}
 & & Bursts: $17-22$ & $69^{\rm o}$ & $129^{\rm o}$ \\
\enddata
\tablenotetext{a}{We calculate the 90\% confidence interval for each burst 
group separately, because this parameter could in principle vary between 
bursts.}
\end{deluxetable}

\clearpage

\begin{deluxetable}{cccc}
\tablecolumns{4}
\tablewidth{0pc}
\tablecaption{Likelihood distribution of the angular radius of the spot 
$\Delta\theta$.}
\tablehead{Burst group\tablenotemark{a} & Parameter & EOS: A18$+\delta v+$UIX & EOS: A18}
\startdata
 & $\Delta\theta$ & Likelihood & Likelihood  \\ \hline
 & $5^{\rm o}$  & $1.000$  & $1.000$  \\
Bursts: $1-8$ & $25^{\rm o}$  & $0.566$ & $0.426$ \\
 & $45^{\rm o}$  & $0.503$ & $0.082$  \\ 
 & $60^{\rm o}$  & $0.047$ & $0.002$  \\
\hline
 & $5^{\rm o}$  & $0.718$  & $0.826$  \\
Bursts: $9-16$ & $25^{\rm o}$  & $0.458$ & $0.712$ \\
 & $45^{\rm o}$  & $1.000$ & $1.000$  \\ 
 & $60^{\rm o}$  & $0.379$ & $0.153$  \\
\hline
 & $5^{\rm o}$  & $0.367$  & $0.569$  \\
Bursts: $17-22$ & $25^{\rm o}$  & $0.439$ & $0.649$ \\
 & $45^{\rm o}$  & $1.000$ & $1.000$  \\ 
 & $60^{\rm o}$  & $0.354$ & $0.110$  \\
\enddata
\tablenotetext{a}{We calculate likelihood distribution for each burst
group separately, because this parameter could in principle vary between
bursts.}
\end{deluxetable}


\end{document}